\documentclass[pra,twocolumn,showpacs,amsmath,amssymb]{revtex4}

\usepackage{graphicx}
\usepackage{dcolumn}
\usepackage{bm}

\begin{document}

\title{Time-of-flight imaging method to observe signatures of antiferromagnetically ordered states of fermionic atoms in an optical lattice
}

\author{Kensuke Inaba}
\author{Makoto Yamashita}%

\affiliation{
NTT Basic Research Laboratories, NTT Corporation, Atsugi 243-0198, Japan
}
\affiliation{
JST, CREST, Chiyoda-ku, Tokyo 102-0075, Japan
}

\date{\today}

\begin{abstract}
We propose a simple method to detect the antiferromagnetic (AF) state of fermionic atoms in an optical lattice by combining a time-of-flight (TOF) imaging method and a Feshbach resonance.
In this scheme, the nontrivial dynamics of fermionic atoms during the imaging process works as a probe with respect to the breaking of the translational symmetry in the AF state.
Precise numerical simulations demonstrate that the characteristic oscillatory dynamics induced by the scattering process that transfers an AF ordering vector appears in TOF images, which can be easily observed experimentally.
\end{abstract}


\preprint{APS/123-QED}

\maketitle

Cold atoms trapped in optical lattices are ideal model systems for investigating the many-body problems that have been studied for many years in condensed matter physics \cite{Greiner2008,Bloch2005}. 
These systems are highly controllable and thus considered to be quantum simulators. 
In fact, the quantum phase transition from a metallic state to a Mott insulating state has been successfully realized using fermionic $^{40}$K atoms in optical lattices \cite{Jordens2008,Schneider2008}. 
The antiferromagnetic (AF) transition in the optical lattice system is now a major concern for condensed matter physicists.
The thermodynamic properties of the AF states in optical lattices have been studied in detail in the recent theoretical work \cite{Snoek2008,Inaba2010}.
On the other hand,  no experimental method has yet been established for detecting such AF states, and this is an important issue.
Several proposals have been reported recently based on the Bragg diffraction of light \cite{Corcovilos2010}, the detection of noise correlations \cite{Bruun2009}, and high-resolution in situ imaging \cite{Bakr2009}, however these methods require extremely sophisticated experimental techniques.
This convinces us of the need for complementary easy-to-use approaches employing conventional experimental methods.

In this paper, we propose a simple method to detect AF states by combining time-of-flight (TOF) imaging and a Feshbach resonance.
A TOF imaging method is a useful standard tool for investigating the optical lattice systems \cite{Pedri2001,Greiner2001}.
Using this method, K\"ohl {\it et al.} observed the Fermi surface of two-component fermionic atoms in an optical lattice, and also observed certain excitations of atoms to higher Bloch bands in combination with a Feshbach resonance \cite{Kohl2005}.
Our present study is highly motivated by this report.
We carry out the precise numerical simulations and then demonstrate that the nontrivial dynamics of fermionic atoms during a TOF imaging process associated with a Feshbach resonance is sensitive to the breaking of the translational symmetry of the system.
We discuss the use of this feature as a probe for the AF state.

We consider a virtual system consisting of fermionic $^{40}$K atoms with two different hyperfine states in an optical lattice.  
We assume that the trapping potential in the vertical direction is much larger than that in the horizontal direction, which effectively makes the system two-dimensional (2D) \cite{Gemelke2009}.
The parameters are set as the relevant values as in the experiments \cite{Kohl2005}; trapping potential frequency $\omega_{\rm trap}=2\pi\times150$\,Hz, scattering length $a_{SC}=1.5$\,nm, lattice laser wavelength $\lambda=826$\,nm, lattice laser beam waist $w=80$\,$\mu$m, and lattice potential depth $V_0=14E_r$, where $E_r=h^2 /(2M\lambda^2)$ is the recoil energy and $M$ is the mass of $^{40}$K atoms.
Furthermore, we consider the balanced population of each hyperfine state.

The present system can be well described by the 2D Hubbard model with a harmonic trapping potential \cite{Zwerger2003}.
The Hamiltonian is written as,
\begin{eqnarray}
{\cal H}&=&\sum_{{\bf r}{\bf d}}\sum_{\sigma\alpha}J_\alpha c^\dag_{{\bf r}\sigma\alpha} c_{{\bf r+d}\sigma\alpha}+\sum_{{\bf r}}\sum_{\alpha\beta} U_{\alpha\beta} n_{{\bf r} \uparrow\alpha} n_{{\bf r} \downarrow\beta}
\nonumber\\
&+&\sum_{\bf r}\sum_{\sigma\alpha} (V_{t\alpha} r^2+\varepsilon_\alpha -\mu) n_{{\bf r}\sigma\alpha}
\label{eq_model},
\end{eqnarray}
where $c^\dag_{{\bf r}\sigma\alpha} (c_{{\bf r}\sigma\alpha})$ creates (annihilates) a fermionic atom of the $\alpha$th Bloch band with pseudospin (hyperfine state) $\sigma(=\uparrow, \downarrow)$ on the lattice site at position ${\bf r}$, and $n_{{\bf r}\sigma\alpha}=c^\dag_{{\bf r}\sigma\alpha}c_{{\bf r}\sigma\alpha}$. 
Here, $J_\alpha$ is the nearest-neighbor hopping integral, $\varepsilon_\alpha$ is the onsite potential, $V_{t\alpha}$ is the curvature of the harmonic trapping potential for atoms in the $\alpha$th band, $U_{\alpha\beta}$ is the on-site interaction between two atoms in the $\alpha$th and the $\beta$th bands with different pseudospins, and $\mu$ is the chemical potential. 
With the exception of $\mu$, these parameters are uniquely calculated from the setup parameters mentioned above, $V_0, \lambda$, and so on.
The chemical potential $\mu$ is determined by fixing the total number of atoms $N$.
The gap between the first and second Bloch bands $\Delta=\varepsilon_2-\varepsilon_1$ is very large in our setup, and the population of the higher bands is very small.
This allows us to carry out the present numerical calculation within the second Bloch bands. 
In the following, we describe for simplicity the quantities of the first Bloch band without indices, for instance, $J_1\equiv J$ and $U_{11}\equiv U$.

There are four energy scales in the present system; temperature $T$, the bandwidth $W=8J$, the interaction strength $U$ and the trapping energy $E_t=V_t N/2\pi$ \cite{Schneider2008}.
Using these scales, we can regard the calculated results as being independent of the details of the parameters, such as $V_t$, $\mu$, and the system size \cite{Schneider2008,Inaba2010}.
Here, we set $W$ as an energy unit.
In Ref. \cite{Inaba2010}, we investigated the AF transition temperature of a 2D optical lattice system. 
We found that the highest transition temperature is obtained at around $U/W\sim1.5$ and $E_t/W\sim0.5$.
In the following calculations, we always realize $U/W\sim1.6$ and $E_t/W\sim0.5$ by choosing the setup parameters. 

Before explaining the details of our calculations, it is useful to summarize how we obtain TOF images of fermions in an optical lattice.
As mentioned in Refs. \cite{Kohl2005}, first the lattice potential is linearly ramped down, namely $V_0$ is linearly decreased to zero.
This ramping down procedure induces nontrivial dynamics, because the parameters of the Hamiltonian (\ref{eq_model}) change as $V_0$ decreases.
Then, the trapping potential is turned off suddenly, and accordingly unconfined atoms expand ballistically in several milliseconds. 
A TOF image of the atoms is finally obtained with a CCD camera.
Such a TOF image ideally corresponds to the momentum distribution of the atoms.
However, the experimentally observed image must be affected by the nontrivial many-body dynamics during the lattice ramping down. 
In addition, TOF images are certainly influenced by the ballistic expansion process in a finite time and the resolution of the CCD cameras \cite{Gerbier2008}.
Our numerical simulations deal with all these effects.

Here we briefly explain procedures of our calculations.
First, we calculate the initial thermal states using a conventional mean-field approach.
The nontrivial dynamics during the lattice ramping down is studied  in the same mean-field manner.
We decompose the ramping time $t_{\rm ramp}$ into small segments characterized by the short time scale $\delta t_{} (=t_{\rm ramp}/L)$.
At each segment, we recalculate the parameters in the Hamiltonian (\ref{eq_model}) according to the time dependence of $V_0$, and simulate the time evolution during $\delta t_{}$ using the exact diagonalization.
We confirm that $L$ is sufficiently large and that this decomposition does not affect the results.
Next we simulate the ballistic expansion of atoms in the time $t_{\rm TOF}$ using the obtained time-evolved states. 
A TOF image is finally evaluated as the spatial distribution of expanded atoms.
We assume that the CCD is a 120$\times$120 array of $2$\,$\mu$m square pixels so that the intensity of a TOF image $I(x,y)$ corresponds to the number of expanded atoms in a 2\,$\mu$m square around the position $(x,y)$. 

\begin{figure}[t]
\includegraphics[width=7.5cm]{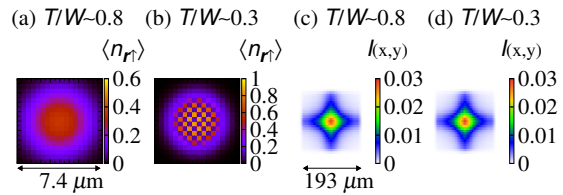}
\caption{(Color Online) 
Number of atoms with $\uparrow$ pseudospin $\langle n_{{\bf r}\uparrow}\rangle$ at (a) $T/W\sim0.8$ and (b) $T/W\sim0.3$.
Time-of-flight image obtained after 2\,ms linear lattice ramping down at (c) $T/W\sim0.8$ and (d) $T/W\sim0.3$.
}
\label{fig_num2d}
\end{figure}
We show the properties of the initial thermal states.
In Fig.\,\ref{fig_num2d}(a) and (b), we show the number of atoms with $\uparrow$ pseudospin for the different temperatures $T/W\sim0.8$ and $T/W\sim0.3$, respectively.
As shown in Fig.\,\ref{fig_num2d}(a), the paramagnetic metallic (PM) state appears at the higher temperature $T/W\sim0.8$.
In Fig.\,\ref{fig_num2d}(b), we find a checker-board pattern around the center of the trapping potential, meaning the AF state appears at the lower temperature $T/W\sim0.3$. 

Next, in Fig.\,\ref{fig_num2d}(c) and (d), we show TOF images calculated for these two different initial states.
Here, we set $t_{\rm ramp}=$2\,ms and $t_{\rm TOF}=$8\,ms.
After 8\,ms of ballistic expansion, the first Brillouin zone (BZ) is converted into a region $|x,y| < 193/2$\,$\mu$m \cite{Greiner2001}.
The population of the higher Bloch bands increases slightly during the lattice ramping down, but remains negligible.
In fact, as shown in Fig.\,\ref{fig_num2d}(c) and (d), there are few signals out of the first BZ.
Although the initial states are quite different from each other, we find no distinctive differences in between the TOF images.
We can understand this feature by noting the following points.
Since a wavevector $\bold{k}$ is not a good quantum number owing to the trapping potential, the difference between the momentum distributions of the AF and PM states is rather subtle even in the initial states.
In addition, the effects of the lattice ramping down make the TOF images featureless, as discussed later. 

\begin{figure}[t]
\includegraphics[width=7.5cm]{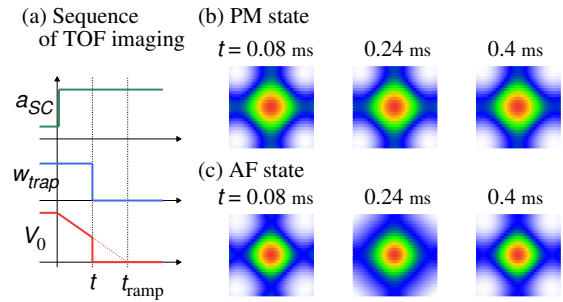}
\caption{(Color Online) (a) Schematic diagram of the sequence of the TOF imaging to detect the AF state.
TOF images taken by the present sequence for (b) the PM state (obtained at $T/W\sim0.8$) and (c) the AF state ($T/W\sim0.3$). 
The image size is $240$\,$\mu$m square.
}
\label{fig_tof2d}
\end{figure}
Here we discuss how to distinguish the AF state from the PM state using TOF images.
For this purpose, we propose the following simple procedure:
(i) before starting the lattice ramping down, we suddenly change the scattering length $a_{SC}$ to $\tilde{a}_{SC}\equiv R_{SC}a_{SC}$ using a Feshbach resonance; 
(ii) then, start the lattice ramping down procedure as usual; 
(iii) at $t (<t_{\rm ramp})$, before the procedure is completed, turn the trapping and lattice potentials off simultaneously; 
(iv) capture a TOF image after ballistic expansion in $t_{\rm TOF}$. 
We depict this sequence in Fig.\,\ref{fig_tof2d}(a).
We repeat sequence (i)-(iv) for different $t$ values, and investigate dynamics of fermionic atoms in the momentum space. 
In the following, we adopt the same PM and AF states as in Fig.\,\ref{fig_num2d} as the initial states, and  set $t_{\rm ramp}=2$\,ms, and $t_{\rm TOF}=8$\,ms.
Although ${R}_{SC}$ is varied up to $10$, the energy scale of the increased interaction strength is much smaller than that of the band gap, $\tilde{U}=R_{SC}U\ll\Delta$. 

In Fig.\,\ref{fig_tof2d}(b) and (c), we show TOF images obtained at $t=0.08, 0.24$, and $0.4$\,ms with $R_{SC}=5$ for the PM and AF states, respectively. 
Even though the population of the higher hands is very small, all of the TOF images in Fig.\,\ref{fig_tof2d}(b) and (c) expand over the 1st BZ because the lattice ramping down is incomplete.
We find that similar cross-shaped images are obtained at $t=0.08$ and $0.40$\,ms for the AF state and at $t=0.08, 0.24$, and $0.40$ for the PM state.
Significantly, a distinctive image is obtained at $t=0.24$\,ms for the AF state.
We find that such distinctive images appear periodically for the AF state from the calculations of a larger $t$.
This kind of characteristic behavior cannot be seen for the TOF images obtained for the PM state.

\begin{figure}[t]
\includegraphics[width=7cm]{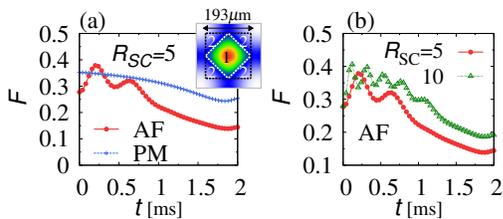}
\caption{(Color Online) (a) and (b) Fraction of atoms in the 2nd RBZ, $F=I_{\rm 2nd}/(I_{\rm 1st}+I_{\rm 2nd})$, as a function of $t$.
The inset shows the definition of the 1st and 2nd RBZs.
}
\label{fig_ovs}
\end{figure}
To clarify this characteristic behavior of the TOF images for the AF state, we introduce a quantity; the fraction of the atoms in the second reduced Brillouin zone (RBZ) $F=I_{\rm 2nd}/(I_{\rm 1st}+I_{\rm 2nd})$, 
where $I_{\rm 1st}$ ($I_{\rm 2nd}$) is a sum of the intensity in the 1st (2nd) RBZ of a TOF image. 
The definitions of the 1st and the 2nd RBZs are given in the inset of Fig.\,\ref{fig_ovs}(a).
The quantity $F$ as a function of $t$ for the PM and AF states with fixed $R_{SC}=5$ is shown in Fig.\,\ref{fig_ovs}(a).
We can clearly see the oscillation of $F$ for the AF state at $t<1$\,ms, while we find no oscillation for the PM state.
At $t>1$\,ms, $F$ gradually decreases for each state.
In Fig.\,\ref{fig_ovs}(b), we show the influence of $R_{SC}$ values on the oscillatory behavior of $F$.
The period of the characteristic oscillation for the AF state becomes faster with increases in ${R_{SC}}$.
The estimated values of the period are $0.44$ and $0.21$\,ms for $R_{SC}=5$ and $10$, respectively.

We can understand this characteristic dynamics of AF states as the signature of translational symmetry breaking.
In our scheme, the scattering length is instantly changed from its initial value via a Feshbach resonance.
This induces some scattering processes as a result of the additional mechanical work imposed on the system. 
The dominant effect comes from the potential scattering ${\cal S}=\tilde{U}\sum_{{\bf r}\sigma} n_{{\bf r}\sigma}\langle n_{{\bf r}\bar{\sigma}}\rangle$.
If the density profile has the checker-board structure shown in Fig.\,\ref{fig_num2d}(b), this potential scattering is rewritten as ${\cal S}\sim \tilde{U}\bar{M}\sum_{{\bf k}\sigma} c^\dag_{{\bf k}\sigma}c_{{\bf k+Q}\sigma}$, where ${\bf Q}=(\pi,\pi)$ is the ordering vector of the AF states, and $\bar{M}$ is an averaged value of the magnetization $M_{\bf r}=1/2|\langle n_{{\bf r}\uparrow}\rangle-\langle n_{{\bf r}\downarrow}\rangle|$.
This scattering, which transfers a wavevector ${\bf Q}$, induces excitations from the 1st RBZ to the 2nd RBZ, resulting in the above mentioned oscillation.
In contrast, for paramagnetic states whose translational symmetry in not broken, ${\cal S}$ does not transfer any wavevector.

\begin{figure}[t]
\includegraphics[width=7cm]{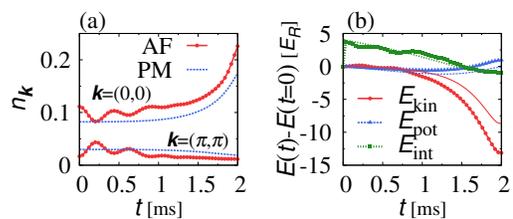}
\caption{(Color Online) (a) Time evolution of the quasi-momentum distribution of atoms with ${\bf k}=(0,0)$ and $(\pi,\pi)$ for the AF and PM states.
(b) Kinetic energy $E_{\rm kin}$, trapping potential energy $E_{\rm pot}$, and interaction energy $E_{\rm int}$ as functions of $t$. Thin lines (thick lines with plots) are data for the PM (AF) state. 
}
\label{fig_ene}
\end{figure}
To discuss this oscillation in more detail, we calculate the quasimomentum distribution $n_{\bf k}$ just before the ballistic expansion. 
In Fig.\,\ref{fig_ene}(a), we show the time evolution of $n_{{\bf k}=(0,0)}$ and $n_{{\bf k}=(\pi,\pi)}$ for the PM and AF states with fixed $R_{SC}=5$.
We find that, at $t<1$\,ms, $n_{{\bf k}=(0,0)}$ and $n_{{\bf k}=(\pi,\pi)}$ for the AF state exhibit an oscillation with the opposite phase, clearly suggesting an excitation transferring ${\bf Q}=(\pi,\pi)$.
This oscillation results from the difference between the energies of quasiparticles with ${\bf k}=(0,0)$ and $(\pi,\pi)$.
In the AF state, the quasiparticle dispersion is approximately given by $\displaystyle \varepsilon_{{\bf k} ({\bf k+Q})}=\frac{1}{2}\left(\epsilon_{0\bf k}+\epsilon_{0\bf k+Q}\pm\sqrt{(\epsilon_{0\bf k}-\epsilon_{0\bf k+Q})^2+4(\bar{M}\tilde{U})^2}\right)$, where $\epsilon_{0\bf k}=-2J\sum_x \cos(k_x)$.
Thus, we obtain the period of this oscillation as $t_{\rm osc}\sim\hbar\pi/\sqrt{(W/2)^2+(\bar{M}\tilde{U})^2}$. 
We obtain $\bar{M}\sim0.3$ from this equation, which is consistent with the value directly calculated from $\langle n_{{\bf r}\sigma}\rangle$ shown in Fig.\,\ref{fig_num2d}(b).
This feature allows us to estimate experimentally the magnetization value via an oscillation period.

In Fig.\,\ref{fig_ene}(a), we find that $n_{{\bf k}=(0,0)}$ for both states increases at $t>1$\,ms, corresponding to the decrease in $F$ shown in Fig.\,\ref{fig_ovs}(a).
To clarify this behavior, in Fig.\,\ref{fig_ene}(b), we show the change in the internal energy during the lattice ramping down for the PM and AF states.
We calculate the kinetic energy $E_{\rm kin}=\sum_{{\bf k}\sigma} \epsilon_{0\bf k} \langle n_{{\bf k}\sigma}\rangle$, trapping potential energy $E_{\rm pot}=V_t \sum_{{\bf r}\sigma} r^2 \langle n_{{\bf r}\sigma}\rangle$, and interaction energy $E_{\rm int}=\tilde{U} \sum_{{\bf r}\sigma}\langle n_{{\bf r}\sigma}\rangle\langle n_{{\bf r}\bar{\sigma}}\rangle$.
The increase in $E_{\rm int}$ in the vicinity of $t=0$\,ms is due to the sudden increase in $a_{sc}$.
We see a large kinetic energy gain for both states, resulting from the fact that $J$ increases exponentially as $V_0$ is linearly decreased \cite{Zwerger2003}.
Note that this effect is more dominant for a larger initial $V_0$ (we now set $V_0=14E_r$).
The large gain of $E_{\rm kin}$ is accompanied by an increase in the number of atoms around the bottom of the first Bloch band $[{\bf k}\sim(0,0)]$, and a decrease in those around the top of the first band $[{\bf k}\sim(\pi,\pi)]$.
This makes the TOF images featureless as mentioned above.

\begin{figure}[t]
\includegraphics[width=7cm]{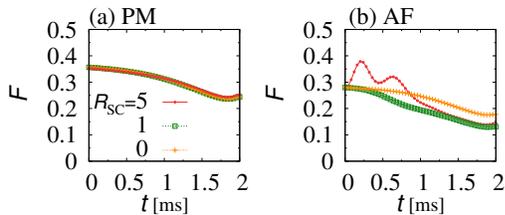}
\caption{(Color Online) Fraction of atoms in the 2nd RBZ $F$ as a function of $t$ for the PM and AF states. The increase rate of the scattering length is changed as $R_{SC}=5, 1,$ and $0$.
}
\label{fig_ovs_as}
\end{figure}
Here, we consider the special cases of $R_{SC}=0$ and $1$, where the oscillation of $F$ disappears even for the initial AF state.
We can easily expect that, without a change in the interaction strength ($R_{SC}=1$), the characteristic oscillation will never occur. 
In addition, when we tune $a_{SC}$ within the noninteracting limit ($R_{SC}=0$), the characteristic oscillation vanishes because the potential scattering ${\cal S}\propto \tilde{U}$ is absent.
In Fig.\,\ref{fig_ovs_as}, we show $F$ for the PM and AF states with $R_{SC}=1$ and $0$, and also that with $R_{SC}=5$ for comparison.
As shown in Fig.\,\ref{fig_ovs_as}(a), the $F$ curves for the PM state change little as $R_{SC}$ is varied.
In contrast, from Fig.\,\ref{fig_ovs_as}(b), we can see the distinctive $F$ curves of the AF state. 
Significantly, we can find that $F$ decreases faster for $R_{SC}=1$ than for $R_{SC}=0$.
This results from the fact that the relaxation that occurs during the lattice ramping down is accelerated by the energy gap of the AF states, $\Delta_{\rm AF}\sim2\tilde{U}\bar{M}$.

Finally, we discuss the behavior of the $F$ curves for the paramagnetic Mott insulating state that we cannot deal with in our mean-field calculation.
Since this state without translational symmetry breaking has an energy gap, we expect to find that $F$ curves show no oscillation but show fast relaxation for $R_{SC}=1$ and $5$, while exhibiting no characteristics for $R_{SC}=0$.


In summary, we investigated a 2D optical lattice system consisting of $^{40}$K atoms with two different hyperfine states.
On the basis of precise numerical simulations of the TOF imaging process with lattice ramping down, we investigated the nontrivial dynamics that occur during this process.
We clarified that the kinetic energy gain induced by the change in the Hamiltonian results in the relaxation from the top to the bottom of the lowest Bloch band.
We also suggested that this relaxation process would be enhanced in the gapped phase.
Furthermore, we proposed a TOF imaging procedure for detecting AF states, where we suddenly change the scattering length using a Feshbach resonance before starting the lattice ramping down and then investigate the dynamical change in the TOF images systematically.
In the translational symmetry broken states, the sudden increase in interaction induces potential scattering, which transfers ordering vector.
This results in the oscillation of the momentum distribution of atoms, which can be detected with the standard TOF imaging technique.
We showed that the signatures of AF states can be detected by the present method.
Note that this method could be applied to other translational symmetry broken phases.
In our calculations, we applied the mean-field approximation to the interaction terms, and discussed only the dominant scattering effects.
Although this might mean that we neglect the higher order scattering process, it is beyond  the scope of our current work.


We thank Y. Takahashi, S. Uetake, S. Taie, S. Sugawa, S. Suga and Y. Tokura for valuable discussions.


\end{document}